\documentclass[%
 reprint,
nofootinbib,
 amsmath,amssymb,
 aps,
]{revtex4-2}
\usepackage{graphicx}
\usepackage{dcolumn}
\usepackage{bm}
\usepackage{enumerate}
\usepackage{amsthm}

\usepackage[dvipsnames]{xcolor}
\usepackage{hyperref}
\hypersetup{
    colorlinks,
    linkcolor={red!50!black},
    citecolor={blue!50!black},
    urlcolor={blue!80!black}
}

\newcommand{\R}{\mathbb R}

\newcommand{\D}{\text{d}}

\makeatletter
\newcommand*{\rom}[1]{\expandafter\@slowromancap\romannumeral #1@}
\makeatother

\newcommand{\SH}[1]{#1}

\begin{document}

\preprint{APS/123-QED}

\title{New class of non-Einstein pp-wave solutions to quadratic gravity}


%


\author{Sjors Heefer}
 \email{s.j.heefer@tue.nl}
\author{Lorens F. Niehof}
 \email{l.f.niehof@student.tue.nl}
\author{Andrea Fuster}
 \email{a.fuster@tue.nl}
\affiliation{%
 Department of Mathematics and Computer Science \\
 Eindhoven University of Technology, Eindhoven 5600MB, The Netherlands
}

\date{\today}

\begin{abstract}
\noindent We obtain a new family of exact vacuum solutions to quadratic gravity that describe pp-waves with two-dimensional wave surfaces that can have any prescribed constant curvature. When the wave surfaces are flat we recover the Peres waves obtained by Madsen, a subset of which forms precisely the vacuum pp-waves of general relativity. If, on the other hand, the wave surfaces have nonvanishing constant curvature then all of our solutions are non-Einstein (i.e. they do not solve Einstein's equations in vacuum, with or without cosmological constant) and we find that the curvature is linearly related to the value of the cosmological constant. We show that the vacuum field equations reduce to a simple linear biharmonic equation on the curved wave surfaces, and as consequence, the general solution can be written down. We also provide some simple explicit examples.
\end{abstract}

\maketitle


\section{\label{sec:introduction} Introduction}
\noindent Quadratic gravity (QG) is a modification of Einstein's general relativity (GR), 
going back to Weyl \cite{Weyl:1918ib}, in which the Einstein-Hilbert action is generalized to contain all possible local terms that are at most quadratic in the curvature and whose coefficients have the dimensionality of non-negative
powers of energy. While we will not be concerned with quantum corrections here, it is worthwhile to point out that, in contrast to the standard perturbative quantization of GR, it was shown already in 1977 by Stelle \cite{Renormalizability_of_quad_grav} that the quadratic gravity action leads to a renormalizable quantum field theory that reduces in the low-energy (low curvature) limit to general relativity.


Einstein spacetimes, i.e. vacuum solutions to general relativity with or without cosmological constant, are trivially solutions in QG. Other than these, however, not many exact solutions to QG are known, due to the complexity of the highly nonlinear fourth-order field equations. Some explicit nontrivial solutions have been obtained in \cite{Madsen_1990, AdS_waves_quad_grav, Malek2011, Gurses_2012,gyratonic_pp_waves_quad_grav,Pravda2016ExactGravity}, all of which belong to the Kundt class or are conformal to it. Moreover, all of these solutions have a constant Ricci scalar, which simplifies the field equations enormously. \SH{Spherically symmetric solutions have been investigated  numerically and by using perturbative methods and series expansions \cite{L__2015, L__20152,Holdom_2017, Podolsky2020BH,Pravda2021}.}

In this paper we introduce a new family of non-Einstein exact pp-wave solutions to QG. We define a pp-wave in the general sense, as a Lorentzian metric that admits a covariantly constant null vector field (CCNV). In general relativity it is then a consequence of Einstein's field equations that (i) the wave surfaces 
of a vacuum pp-wave are always flat in standard coordinates, and that (ii) pp-waves are not reconcilable with a nonvanishing cosmological constant\footnote{We note, however, that the related Siklos solutions, which are (in some sense) conformal to pp-waves, can be interpreted as exact gravitational waves traveling on an AdS background \cite{Podolsky_1998}. Their generalization to QG has been studied in \cite{AdS_waves_quad_grav}.}.  The solutions that we will present here, on the other hand, show that QG allows for vacuum pp-waves with wave surfaces of any constant curvature, 
where this curvature is proportional to the value of the cosmological constant. We show that the field equations reduce to a simple linear biharmonic equation on the wave surfaces, for which the general solution can be written down. When the wave surfaces are assumed to be flat, we recover the classical pp-waves of GR and their QG generalizations \cite{Madsen_1990}.



\section{\label{sec:QG} QG Action and Field Equations}
\noindent The action for quadratic gravity contains, by definition, all possible local terms that are at most quadratic in the curvature and whose coefficients have the dimensionality of non-negative powers of energy, that is, terms proportional to $R, R^2, R_{\mu\nu}R^{\mu\nu}$, and $R_{\mu\nu\rho\sigma}R^{\mu\nu\rho\sigma}$, 
as well as a possible cosmological constant term. We will use the following conventions for the curvature tensors and scalar,
\begin{align}
    R^\rho{}_{\sigma\mu\nu} &= \partial_\mu\Gamma^\rho_{\nu \sigma} - \partial_\nu\Gamma^\rho_{\mu \sigma} + \Gamma^\rho_{\lambda \mu}\Gamma^\lambda_{\nu\sigma }  - \Gamma^\rho_{\lambda\nu}\Gamma^\lambda_{\mu\sigma},\\
    R_{\mu\nu} &= R^\rho{}_{\mu\rho\nu},\\
     R &= R^\mu{}_\mu
\end{align}
as well as signature convention $(-+++)$. In four spacetime dimensions, it follows from the Chern-Gauss-Bonnet theorem and the Ricci decomposition of the curvature tensor that the $R_{\mu\nu\rho\sigma}R^{\mu\nu\rho\sigma}$ term and the $R_{\mu\nu}R^{\mu\nu}$ term can simultaneously be eliminated in favor of a single $C_{\mu\nu\rho\sigma}C^{\mu\nu\rho\sigma}$ term involving the Weyl tensor,
\begin{align}
    C_{\mu\nu\rho\sigma} = &\,\,R_{\mu\nu\rho\sigma } \\
    &+ \frac{1}{2}\left(g_{\mu \sigma }R_{\nu \rho } + g_{\nu \rho }R_{\mu \sigma } - g_{\mu \rho }R_{\nu \sigma } - g_{\nu \sigma }R_{\mu \rho } \right) \nonumber\\
    &+ \frac{R}{6} \left(g_{\mu \rho } g_{\sigma \nu } - g_{\mu \sigma } g_{\rho \nu } \right). \nonumber
\end{align}
One thus arrives at the following action,
\begin{align}
    %
    %
    S =\!\! \int\!\! \D^4 x\sqrt{-g} \left(\gamma'(R-2\Lambda) + \alpha'C_{\mu\nu\rho\sigma}C^{\mu\nu\rho\sigma} \!+ \!\beta' R^2\right)\!,
\end{align}
where $\alpha', \beta', \gamma'$ are constants. The vacuum field equations obtained by extremizing this action are given by
\begin{align}\label{eq:FEQG}
    &\gamma'\left(R_{\mu\nu} - \frac{1}{2}Rg_{\mu\nu} + \Lambda g_{\mu\nu} \right) - 4 \alpha' B_{\mu\nu} \nonumber \\
    &+ 2\beta'\left(R_{\mu\nu} - \frac{1}{4}R g_{\mu\nu} + g_{\mu\nu} \Box - \nabla_\nu \nabla_\mu \right)R = 0,
\end{align}
where  
\begin{align}\label{eq:box_operator_definition}
    \Box =  g^{\mu\nu} \nabla_\mu \nabla_\nu
\end{align}
is the covariant d'Alembertian 
and $B_{\mu\nu}$ is the Bach tensor, defined as
\begin{align}\label{eq:bachtensor0}
    B_{\mu\nu} &= \left(\nabla^\rho\nabla^\sigma + \frac{1}{2}R^{\rho\sigma}\right)C_{\mu\nu\rho\sigma},
\end{align}
which can also be written as\footnote{To avoid confusion, we stress that in the expression $\Box R_{\mu\nu}$, the d'Alembertian acts on the tensor $R_{\mu\nu}$ \textit{as a whole}, not on the component functions $R_{\mu\nu}$ individually, which is sometimes the case in other applications. 
}
\begin{align}\label{eq:bachtensor}
    B_{\mu\nu} &= \frac{1}{2} \Box R_{\mu\nu} - \frac{1}{6}\left(\nabla_\mu \nabla_\nu + \frac{1}{2}g_{\mu\nu} \Box \right) R - \frac{1}{3}R R_{\mu\nu} \nonumber \\
    &+ R_{\mu\nu\rho\sigma}R^{\rho\sigma} + \frac{1}{4}\left(\frac{1}{3}R^2 - R_{\rho\sigma}R^{\rho\sigma} \right)g_{\mu\nu}.
\end{align}

\noindent From the second expression, it can be inferred that the Bach tensor vanishes identically for Einstein spacetimes, as these satisfy
\begin{align}\label{eq:EinsteinSpace}
    R_{\mu\nu} = \frac{1}{4}R g_{\mu\nu}
\end{align}
with $R =$ const. Similarly, it is easy to see that the whole second line in \eqref{eq:FEQG} vanishes in that case as well. In other words, any vacuum solution to Einstein gravity (with or without cosmological constant) is automatically a vacuum solution to QG as well.

It is notoriously difficult, on the other hand, to find exact solutions to \eqref{eq:FEQG} that are not Einstein. The field equations nevertheless simplify considerably in the case that the scalar curvature $R$ is constant (note the latter does not imply that the space is Einstein). To see this, note that in that case the trace of \eqref{eq:FEQG} becomes
\begin{align}
    \gamma'(-R+4\Lambda) = 4\alpha' g^{\mu\nu}B_{\mu\nu} = 0,
\end{align}
where we have used the fact that the Bach tensor is traceless. With $ \gamma'\neq 0$ this implies that $R = 4 \Lambda$, and hence the vacuum field equations \eqref{eq:FEQG} reduce to
\begin{align}\label{eq:reducedQGE}
    \left(\gamma' + 8 \beta' \Lambda \right) \left(R_{\mu\nu} - \Lambda g_{\mu\nu}\right) = 4\alpha' B_{\mu\nu}.
\end{align}

\noindent In the specific case that $\Lambda = -\frac{\gamma'}{8 \beta'}$, it is clear that metrics with constant scalar curvature and vanishing Bach tensor provide solutions to QG. Such scenarios have been discussed in \cite{Pravda2016ExactGravity}. Here we will assume that $\Lambda \neq -\frac{\gamma'}{8 \beta'}$. In this case, Equation \eqref{eq:reducedQGE} can be reformulated as
\begin{align}\label{eq:simplifiedFEQG}
    R_{\mu\nu} - \Lambda g_{\mu\nu} - 4\kappa B_{\mu\nu} = 0,
\end{align}
with 
\begin{align}\label{eq:def_kappa}
    \kappa = \frac{\alpha'}{\gamma' + 8\beta'\Lambda}.
\end{align}

\section{\label{sec:pp_waves_definition} pp-Waves}

\noindent The solutions that we will present in this paper belong to the class of pp-waves (plane-fronted waves with parallel rays). In general, pp-waves are Lorentzian metrics defined by the property that they admit a CCNV. They were described for the first time by Brinkmann \cite{Brinkmann:1925fr} and have been studied in detail in e.g. \cite{EhlersKundt1960,ehlers1962exact} in the context of GR (see section 24.5 in \cite{stephani_kramer_maccallum_hoenselaers_herlt_2003} for a summary). For an algebraic classification independent of any field equations, see \cite{Podolsky_2013}, and see \cite{Roche2023} for an enlightening recent review. Roughly speaking pp-waves can be interpreted as exact gravitational waves traveling on a Minkowski background.

It is well known that a pp-wave metric can always be expressed in suitable local coordinates as
\begin{align}
\D s^2 &=  2\D u \left(\D v + H(u,x)\, \D u + \,W_a(u,x)\,\D x^a\right) \label{eq:original_pp_wave_metric}\\
&\qquad\qquad\qquad\qquad\qquad\quad+h_{ab}(u,x) \D x^a \D x^b, \nonumber
\end{align}
where $H, W_2, W_3$ are smooth functions and $h_{ab}$ is a $u$-dependent smooth family of Riemannian metrics of dimension two on the wave surfaces. Here and throughout the remainder of the article we will adhere to the convention that greek indices $\mu,\nu,\dots$ run from $0$ to $3$, while latin indices $a,b,\dots$ run from $2$ to $3$.

Whether the pp-wave metric \eqref{eq:original_pp_wave_metric} can in general be simplified further depends on the specific theory that one is interested in. For example, in general relativity, it can be shown that for any vacuum pp-wave (i.e. a metric of the form \eqref{eq:original_pp_wave_metric} that solves Einstein's field equations in vacuum) the metric functions $W_a$ can be eliminated and $h_{ab}$ may be chosen as $\delta_{ab}$, by a suitable coordinate transformation. The metric then takes the simple form
\begin{align}
\D s^2 = 2\D u \left(\D v + H(u,x)\, \D u\right) +\delta_{ab} \D x^a \D x^b,\label{eq:reduced_pp_wave_metric}
\end{align}
in so-called Brinkmann coordinates. Some authors even take this expression for the metric to be the definition of a pp-wave. In the context of vacuum GR this can be defended, but in general, and in particular in quadratic gravity, pp-waves need not have the simple form \eqref{eq:reduced_pp_wave_metric}.

We define the wave surfaces of a given pp-wave metric as the two-dimensional Riemannian submanifolds of spacetime given by $\D u = \D v = 0$.\footnote{We note that some authors choose to define the wave surfaces as the three-dimensional hypersurfaces given by $\D u=0$ rather than the two-dimensional surfaces $\D u=\D v=0$. } While in GR these may always be assumed to be flat, we will see below that this is not the case in QG.

\section{\label{sec:exactsolutions} Vacuum pp-waves in quadratic gravity}

\noindent Rather than having the ambition to classify all possible vacuum pp-waves in QG, our point of departure will be the following ansatz,

\begin{align}\label{eq:metric_ansatz}
    ds^2 &= 2 H(u,x,y)du^2 + 2du dv  + \D \tilde s^2,
\end{align}
where $\D \tilde s^2$ is a two-dimensional metric characterizing the geometry of the wave surfaces, which we will assume to be of constant curvature\footnote{Hence the wave surfaces are, in particular, Einstein as well.} Then \eqref{eq:metric_ansatz} belongs in particular to the class of CSI spacetimes\footnote{These are spacetimes with constant scalar curvature invariants.} \cite{Coley_2009}. We thus consider the following three possibilities:
\begin{enumerate}[A.]
    \item the wave surfaces are flat;
    \item the wave surfaces have constant \textit{positive} curvature, i.e. are spherical;
    \item the wave surfaces have constant \textit{negative} curvature, i.e. have hyperbolic geometry.
\end{enumerate}
Clearly this ansatz falls under the general definition of a pp-wave given above. 
However, whereas the classes of metrics $(B)$ and $(C)$ can never be vacuum solutions to Einstein's equations, as we will confirm explicitly below, we will find that they do solve the vacuum equations of QG, provided that the function $H$ satisfies the differential equation $\Delta^2 H = 0$, where $\Delta$ is the 2D Laplace-Beltrami operator on the wave surfaces. \SH{Interestingly, related GR solutions with nonvanishing cosmological constant do exist. These represent impulsive gravitational waves generated by null point sources, and have wave surfaces of constant curvature \cite{Hotta1993,Podolsky1997}.}

Metrics in class $(A)$, on the other hand, are precisely metrics of the form \eqref{eq:reduced_pp_wave_metric}, and it is well known that there is a subset of this class that solves Einstein's field equations in vacuum, characterized by $\delta^{ab}\partial_a\partial_b H =0$. Nevertheless, in QG the class of solutions of this form is larger. This case was already covered in \cite{Madsen_1990} and we will only review it briefly below.

Before we move on to the analysis of the field equations, it is worth pointing out that our ansatz is in some sense orthogonal to the  gyratonic pp-wave ansatz recently analyzed in \cite{gyratonic_pp_waves_quad_grav} for quadratic gravity, where the appropriate generalization of \eqref{eq:reduced_pp_wave_metric} is the presence of nonvanishing $W_a$ as in \eqref{eq:original_pp_wave_metric}. Our present approach, on the other hand, is to stick to $W_a=0$, while instead generalizing the flat $\delta_{ab}$ in \eqref{eq:reduced_pp_wave_metric} to a curved $h_{ab}$ as in \eqref{eq:original_pp_wave_metric}.


\subsection{\label{sec:Brinkmann} Flat wave surfaces}

\noindent We start with the case of flat wave surfaces, leading to a standard Brinkmann metric
\begin{align}\label{eq:Brinkmannmetric0}
    ds^2 = 2 H(u,x,y) du^2 + 2du dv +  dx^2 + dy^2,
\end{align}
where $H$ is an arbitrary smooth function. The only nonvanishing component of the Ricci tensor is given by
\begin{align}\label{eq:ric00}
    R_{uu} &= -\left(\frac{\partial^2 H}{\partial x^2} + \frac{\partial^2 H}{\partial y^2} \right).
\end{align}
and the only nonvanishing component of the Bach tensor is given by
\begin{align}\label{eq:B00}
    B_{uu} &= -\frac{1}{2}\Delta_{\mathbb E}^2 H.
    %
\end{align}
Here we have denoted by
\begin{align}
    \Delta_{\mathbb E}  \equiv \frac{\partial^2 }{\partial x^2} + \frac{\partial^2 }{\partial y^2}
\end{align}
the two-dimensional Euclidean Laplace-Beltrami operator on the wave surfaces. The Laplace-Beltrami operator corresponding to a general (pseudo-)Riemannian manifold is defined as
\begin{align}
    \Delta f &= g^{ab}\nabla_a\nabla_b f \nonumber \\
    &= g^{ab} \partial_a \partial_b f - g^{bc} \Gamma^{a}_{bc} \partial_i f \label{eq:Laplace_Beltrami_general}\\
    &= \frac{1}{\sqrt{|g|}}\partial_a\left(\sqrt{|g|} g^{ab}\partial_b f \right), \nonumber
\end{align}
where $f$ is an arbitrary function and $g=\det g_{ab}$. The Laplace-Beltrami operator formally coincides with the d'Alembertian \eqref{eq:box_operator_definition}, but throughout this paper we will reserve $\Delta$ for the (Riemannian) wave surfaces, whereas $\Box$ will always be understood to represent the d'Alembertian of the full four-dimensional spacetime metric. The subscript $\mathbb E$ in $\Delta_{\mathbb E}$ serves to distinguish the Euclidean Laplace-Beltrami operator from the spherical and hyperbolic ones that we will encounter later on. 

Plugging \eqref{eq:ric00} and \eqref{eq:B00} into the field equations \eqref{eq:simplifiedFEQG}, 
the $uv$-component immediately implies that $\Lambda = 0$, and consequently the only other nontrivial component takes the form
\begin{align}
     \boxed{\left(2 \kappa \Delta_{\mathbb E}^2 - \Delta_{\mathbb E}\right) H =0,}
\end{align}
where we recall that $\kappa$ is defined as in \eqref{eq:def_kappa}. This agrees\footnote{Our $\kappa$ is related to $K$ in \cite{Madsen_1990} by $K^2 = -1/(2\kappa)$.} with the results in \cite{Madsen_1990}. The general solution to this equation has been constructed in \cite{Madsen_1990} as well and there is no need to repeat that analysis here.

Finally, we note as a consequence of \eqref{eq:ric00} that \eqref{eq:Brinkmannmetric0} is an Einstein metric if and only if $\Delta_{\mathbb E} H  = 0$ (and in this case it also follows that $R = \Lambda = 0$). Hence pp-waves with flat wave surfaces are less constrained in quadratic gravity than they are in general relativity. As we will see below, this is even more true for pp-waves with curved wave surfaces.

\subsection{\label{sec:positive_curvature_pp_waves} Spherical wave surfaces}

\subsubsection{The field equations}

\noindent Next, we turn to wave surfaces of constant positive curvature. Without loss of generality, we may choose coordinates such that the pp-wave metric has the form
\begin{align}\label{eq:4DMetricInteresting0}
    d&s^2 = \nonumber\\
    &2 H(u,\theta,\phi)du^2 + 2du dv + r^2 \left(d\theta^2 + \sin^2{\theta}\, d\phi^2 \right).
\end{align}

%
%
\noindent The nonvanishing components of the Ricci tensor are then given by:
\begin{align}
    R_{\theta\theta} &= 1, \\
    R_{\phi \phi } &= \sin^2{\theta}, \\
    R_{uu} &= - \frac{1}{r^2}\Delta_{\mathbb S} H,
\end{align}
where 
\begin{align}
    \Delta_{\mathbb S} =  \frac{1}{\sin{\theta}}\frac{\partial}{\partial \theta} \left(\sin \theta \frac{\partial}{\partial \theta}  \right) + \frac{1}{\sin^2{\theta}}\frac{\partial^2}{\partial \phi^2}
\end{align}
is the Laplace-Beltami operator \eqref{eq:Laplace_Beltrami_general} on the (unit) 2-sphere. From this, we find that the Ricci scalar satisfies
\begin{align}
    R = \frac{2}{r^2},
\end{align}
and hence the requirement that $R = 4\Lambda$ yields
\begin{align}
    \Lambda = \frac{1}{2 r^2}.
\end{align}
In particular, the cosmological constant must be \textit{positive}. Note that the metric \eqref{eq:4DMetricInteresting0} is never Einstein, as can be seen, for instance, from the fact that
\begin{align}
    R_{\theta\theta} - \frac{1}{4} R g_{\theta\theta} &= \frac{1}{2} \neq 0,
\end{align}
showing that \eqref{eq:EinsteinSpace} is not satisfied. The nonzero components of the Bach tensor are given by
\begin{align}
    B_{uu} 
    &= - \frac{1}{r^4}\left(\frac{1}{2} \Delta_{\mathbb S}^2 + \frac{1}{3 } \Delta_{\mathbb S} + \frac{1}{3 } \right) H  \\
    B_{uv} &= B_{vu} = -\frac{1}{6 r^4},  \\
    B_{\theta\theta} &= \frac{1}{6 r^2}, \qquad   B_{\phi \phi } = \frac{\sin^2{\theta}}{6 r^2}.
\end{align}
Plugging this into the field equations \eqref{eq:simplifiedFEQG}, we find that the $uv, \theta\theta$ and $\phi \phi $ components,
\begin{align}
    0 &= R_{uv} - \Lambda g_{uv} - 4 \kappa B_{uv} = \frac{4\kappa - 3r^2}{6r^4}, \\
    0 &= R_{\theta\theta} - \Lambda g_{\theta\theta} - 4 \kappa B_{\theta\theta} = \frac{3r^2 - 4\kappa}{6r^2}, \\
    0 &= R_{\phi \phi } - \Lambda g_{\phi \phi } - 4 \kappa B_{\phi \phi } = \frac{3r^2 - 4\kappa}{6r^2} \sin^2{\theta},
\end{align}
all lead to the same constraint, namely that
\begin{align}\label{eq:cond1}
    4\kappa = 3 r^2 = \frac{3}{2\Lambda},
\end{align}
or, using the definition \eqref{eq:def_kappa} of $\kappa$, that
\begin{align}
    \alpha' = \frac{3}{8\Lambda}\left(\gamma' + 8 \beta' \Lambda\right).
\end{align}
Since $\left(\gamma' + 8 \beta' \Lambda\right)\neq 0$, by assumption, it follows that $\alpha'\neq 0$, and hence that
\begin{align}
    8\Lambda\left(\alpha' - 3\beta'\right) = 3\gamma'.
\end{align}
Assuming that $\gamma'\neq 0$ as well (which is needed for consistency with GR in the low-energy limit) it follows that $\alpha' - 3\beta'\neq 0$, and hence we may rewrite the relation as
\begin{align}\label{eq:spherical_cosm_const_value}
    \Lambda = \frac{3\gamma'}{8\left(\alpha' - 3\beta'\right)}.
\end{align}
This is an interesting result, as it means that for a given choice of the coefficients $\alpha', \beta'$ and $\gamma'$, pp-waves with spherical wave surfaces may exist only for (at most) a single value of the cosmological constant, namely the one given by \eqref{eq:spherical_cosm_const_value}. Since $\Lambda = 1/2r^2>0$, the right-hand side of \eqref{eq:spherical_cosm_const_value} must in particular be positive for these solutions to exist.

So let us assume, throughout the remainder of this section, that $\alpha', \beta', \gamma'$ and $\Lambda$ are related as above. The only remaining nontrivial component of the field equations \eqref{eq:simplifiedFEQG} reads
\begin{align}
    R_{uu} - \Lambda g_{uu} - 4\kappa B_{uu} = \frac{3}{2r^2}\Delta_{\mathbb S}^2 H = 0,
\end{align}
so that the metric \eqref{eq:4DMetricInteresting0} is an exact solution if and only if $H$ satisfies the biharmonic equation
\begin{align}\label{eq:PDEonH0}
    \boxed{\Delta_{\mathbb S}^2 H = 0.}
\end{align}
In other words, the mapping $(\theta,\phi )\mapsto H(u,\theta,\phi )$ must be a biharmonic function on the sphere. %
%
%
It is worth pointing out that, because $H$ does not depend on $v$, the action of the Laplace-Beltrami operator $\Delta_{\mathbb S}$ is proportional to action of the four-dimensional d'Alembertian \eqref{eq:box_operator_definition}. More precisely, we have $r^2\Box = \Delta_{\mathbb S}$. Hence we may equivalently think of the statement above as saying that a pp-wave with spherical wave surfaces \eqref{eq:4DMetricInteresting0} is a vacuum solution if and only if $H$ satisfies
\begin{align}
    \Box^2 H = 0,
\end{align}
which may be thought of as a kind of generalized wave equation. 

\subsubsection{Globally spherical solutions to $\Delta_{\mathbb S}^2 H = 0$}

\noindent In order to find the general solution, we return to the formulation in \eqref{eq:PDEonH0}. It is a well-known result that there are no harmonic functions on the sphere, other than the constant ones, so one might wonder if the class of (smooth) biharmonic functions is any larger. This turns out not to be the case. We will prove momentarily that any biharmonic function on the sphere is necessarily constant. This assumes, however, that the wave surfaces are \textit{globally} spherical. In contrast, we will show in Section \ref{sec:solving1} below that, locally, nontrivial solutions do exist. We construct the general solution and provide a simple example.

Within the class of square integrable (real-valued) functions $L^2(\mathbb S^2)$ (which includes all smooth functions, since $\mathbb S^2$ is compact), the general solution can be found using an expansion in spherical harmonics. These are the eigenfunctions $Y_{\ell m}$ of $\Delta_{\mathbb S}$, where $\ell=0,1,2,\dots$ and $m = -\ell, -\ell+1,\dots,\ell-1,\ell$, that satisfy 
\begin{align}\label{eq:eigenfunction_eq_spherical_harmonics}
     \Delta_{\mathbb S} Y_{\ell m} = -\ell(\ell+1)Y_{\ell m}.
\end{align}
They form an orthonormal basis of the Hilbert space $L^2(\mathbb S^2)$, which means that we can expand 
\begin{align}
    H = \sum_{\ell=0}^\infty\sum_{m=-\ell}^\ell H_{\ell m} Y_{\ell m}
\end{align}
with $H_{\ell m}\in\R$, and consequently
\begin{align}
    0 &= \Delta_{\mathbb S}^2H \\
    &= \sum_{\ell=0}^\infty\sum_{m=-\ell}^\ell H_{\ell m} \Delta_{\mathbb S}^2 Y_{\ell m} \label{eq:term-by-term_differentiation} \\
    &=  \sum_{\ell=0}^\infty\sum_{m=-\ell}^\ell H_{\ell m} \ell^2(\ell+1)^2 Y_{\ell m}.
\end{align}
The term-by-term differentiation is justified by virtue of the fact that $H$ is smooth (see Appendix \ref{app:spherical_harmonics_expansion} for a proof sketch). Since the $Y_{\ell m}$ form an orthonormal basis, this can only be satisfied if all the coefficients vanish, $H_{\ell m} \ell^2(\ell+1)^2=0$. It follows therefore that $H_{\ell m}=0$ whenever $\ell\neq 0$, and hence that $H = H_{00}Y_{00} = $ constant.

This analysis concerns $H(u,\theta,\phi )$ as a function of $\theta$ and $\phi $, i.e. for each fixed value of $u$. In other words, the result shows that we must have $H = H(u)$ and hence the line element \eqref{eq:4DMetricInteresting0} reduces to
\begin{align}
    ds^2 &= 2 H(u)\D u^2 + 2\D u \D v + r^2 \left(\D\theta^2 + \sin^2{\theta} \,\D\phi ^2 \right),
\end{align}
which is a direct product of a two-dimensional metric with coordinates $(u,v)$ and a 2-sphere of radius $r$. The former 2D metric turns out to have a vanishing Riemann tensor, meaning that there exists a coordinate transformation $(u,v)\mapsto (t,z)$ such that 
\begin{align}
    2 H(u)\D u^2 + 2\D u \D v = -\D t^2 + \D z^2.
\end{align}
It follows that the only vacuum pp-wave in QG with (globally) spherical wave surfaces is therefore given by the static metric
\begin{align}
    ds^2 = -\D t^2 + \D z^2  + r^2 \left(\D\theta^2 + \sin^2\theta\, \D\phi^2 \right).
\end{align}

\subsubsection{\label{sec:solving1} Local solutions to $\Delta_{\mathbb S}^2 H = 0$}

\noindent If, alternatively, we interpret the line element only locally, in other words not enforcing that the wave surfaces should be globally spherical, then in many cases an explicit formula can be given for the general solution. To this end, we will find it convenient to work in stereographic coordinates $(x,y)$, that are related to $(\theta,\phi)$ in the following way,

\begin{align}
    x = \frac{\sin\theta\cos\phi}{1-\cos\theta}, \qquad y = \frac{\sin\theta\sin\phi}{1-\cos\theta}.
\end{align}
In these coordinates, which cover the whole sphere except for the north pole $\theta=0$, the metric on the wave surface attains the form
\begin{align}\label{eq:stereographic_metric}
    \D s^2 = \frac{4r^2\left(\D x^2 + \D y^2\right)}{(1+x^2+y^2)^2},\qquad x,y\in\R,
\end{align}
and the Laplace-Beltrami operater can be written as
\begin{align}
    \Delta_{\mathbb S} H= f(x,y)\Delta_{\mathbb E} H,
\end{align}
where $\Delta_{\mathbb E} = \partial_x^2  + \partial_y^2$ is the Euclidean Laplace-Beltrami operator, and 
\begin{align}
    f(x,y)=\tfrac{1}{4r^2}\left(1+x^2 + y^2\right)^2\neq 0.
\end{align}
In these coordinates, harmonic functions on $\mathbb S^2\setminus\{N\}$, where $N$ is the north pole, are therefore given simply by harmonic functions on $\mathbb R^2$. The removal of a single point thus already makes a big difference for the space of solutions. But in fact, since our present philosophy is that the line element only locally has the specified form \eqref{eq:4DMetricInteresting0}, and that the wave surfaces are only locally given by \eqref{eq:stereographic_metric}, we really are interested in solutions on an arbitrary open subset $U\subset\R^2$. Now consider the equation $\Delta_{\mathbb S}^2 H=0$. It is clearly equivalent to the pair of equations
\begin{align}
    \Delta_{\mathbb S}H = \phi, \qquad \Delta_{\mathbb S}\phi = 0,
\end{align}
which in turn is equivalent, using the form of $\Delta _{\mathbb S}$, to
\begin{align}
    \Delta_{\mathbb E}H &= \frac{\phi}{f}, \label{eq:effective_Poisson_eq0}\\
    \Delta_{\mathbb E}\phi &= 0.\label{eq:effective_Laplace_eq0}
\end{align}
Thus, one first has to solve the standard Laplace equation \eqref{eq:effective_Laplace_eq0} on $U$, and then the Poisson equation \eqref{eq:effective_Poisson_eq0}. There are standard techniques for doing this (see e.g. \cite{hackbusch2017elliptic} or any standard textbook on partial differential equations), but since $U$ is arbitrary we do not consider it worthwhile at this point to specify a possible general solution any further. Instead, we give a simple example below. \\
%
%
%

\subsubsection{A simple example}

\noindent A simple example of an exact solution with $H = H(\theta)$ can be given in our original spherical coordinates \eqref{eq:4DMetricInteresting0}. Consider, for instance, the real-valued function
\begin{align}\label{eq:conditionsinteresting}
H(\theta) &=\frac{1}{2}\ln\left(\frac{1+\cos\theta}{1-\cos\theta}\right).
\end{align}
Because of the branch cut of the natural logarithm, this function cannot be extended to a smooth function on the whole sphere, but for $\theta\in (0,\pi)$, say, it is harmonic and hence in particular biharmonic, i.e. it satisfies the field equations \eqref{eq:PDEonH0}. Hence, for $r$ given by \eqref{eq:cond1}, this example yields an exact vacuum solution to the field equations of QG with a value of the cosmological constant given by \eqref{eq:spherical_cosm_const_value}. Note that it is \textit{not} a vacuum solution to Einstein's equations, since the metric is not Einstein, as shown above.


We end this section by pointing out that the class of biharmonic functions on open subsets of the sphere, and hence the space of vacuum pp-wave solutions with (locally) spherical wave surfaces, is strictly larger than the class of harmonic functions, as illustrated by the proper biharmonic functions constructed in \cite{GUDMUNDSSON2018244}.

\subsection{\label{sec:negative_curvature_pp_waves} Hyperbolic wave surfaces}

\subsubsection{The field equations}\label{sec:hyp_field_eqs}

\noindent Next, we turn to wave surfaces of constant negative curvature. For this we employ the Poincar\'e half-space model for the hyperbolic plane, which may be defined as the manifold $\{(x,y)\in\R\,:\, x>0\}$ endowed with the metric $\D s^2 = \left(dx^2 + dy^2 \right)/x^2$. Rescaling this metric by $r^2$ yields a space of Gaussian curvature $-|r|$. The complete spacetime metric thus becomes
\begin{align}\label{eq:4DMetricHyperbolicInteresting0}
    ds^2 = 2 H(u,x,y)du^2 + 2du dv + \frac{r^2}{x^2} \left(dx^2 + dy^2 \right),
\end{align}
where $H$ is again an arbitrary smooth function. The nonzero components of the Ricci tensor are then given by:
\begin{align}
    R_{xx} &=  R_{yy}=-\frac{1}{x^2}, \\
    R_{uu} 
    &= - \frac{1}{r^2}\Delta_{\mathbb H} H,
\end{align}
where
\begin{align}
   \Delta_{\mathbb H} = x^2 \left(\frac{\partial^2 }{\partial x^2} + \frac{\partial^2 }{\partial y^2}\right)
\end{align}
is the Laplace-Beltami operator \eqref{eq:Laplace_Beltrami_general} in the standard hyperbolic plane (i.e. with Gaussian curvature -1). The Ricci scalar is given by
\begin{align}
    R = -\frac{2}{r^2},
\end{align}
and hence the requirement that $R = 4\Lambda$ yields
\begin{align}
    \Lambda = -\frac{1}{2 r^2}.
\end{align}
In particular, the cosmological constant must be \textit{negative}. Note that the metric \eqref{eq:4DMetricHyperbolicInteresting0} is not Einstein, regardless of the choice of $H$, which can be seen, for instance, from the fact that
\begin{align}
    R_{xx} - \frac{1}{4} R g_{xx} &= - \frac{1}{2 x^2} \neq 0, 
\end{align}
showing that \eqref{eq:EinsteinSpace} is not satisfied. The nonzero components of the Bach tensor are given by
\begin{align}
    B_{uu} &= - \frac{1}{r^4}\left[\frac{1}{2} \Delta_{\mathbb H}^2 -\frac{1}{3} \Delta_{\mathbb H} + \frac{1}{3} \right] H,\\
    B_{uv} &= B_{vu} = -\frac{1}{6 r^4}, \\
    B_{xx} &= B_{yy}=\frac{1}{6 r^2 x^2}, 
\end{align}
Plugging this into the field equations \eqref{eq:simplifiedFEQG}, we find that the $uv, xx$ and $yy$ components,
\begin{align}
    0 &= R_{uv} - \Lambda g_{uv} - 4 \kappa B_{uv} = \frac{4\kappa + 3r^2}{6r^4}, \\
    0 &= R_{xx} - \Lambda g_{xx} - 4 \kappa B_{xx} = -\frac{4\kappa + 3r^2}{6r^2 x^2}, \\
    0 &= R_{yy} - \Lambda g_{yy} - 4 \kappa B_{yy} = -\frac{4\kappa + 3r^2}{6r^2 x^2}.
\end{align}
all lead to the same constraint, namely that
\begin{align}\label{eq:cond1hyperbolic}
    4\kappa = -3 r^2 = \frac{3}{2\Lambda}.
\end{align}
This leads to the same conditions on the parameters in the QG Lagrangian as the ones discussed for the spherical wave surfaces right below Eq. \eqref{eq:cond1}, except that in this case the value of \eqref{eq:spherical_cosm_const_value} must be negative rather than positive. More precisely, we find again that
\begin{align}\label{eq:spherical_cosm_const_value2}
    \Lambda = \frac{3\gamma'}{8\left(\alpha' - 3\beta'\right)}
\end{align}
Just as in the case with spherical wave surfaces, this means that for a given choice of the coefficients $\alpha', \beta'$ and $\gamma'$, pp-waves with hyperbolic wave surfaces may exist only for (at most) a single value of the cosmological constant, namely the one given by \eqref{eq:spherical_cosm_const_value2}. Since $\Lambda = -1/2r^2<0$, the right-hand side of \eqref{eq:spherical_cosm_const_value2} must in particular be negative for such solutions to exist.

So let us assume, throughout the remainder of this section, that $\alpha', \beta', \gamma'$ and $\Lambda$ are related as above. The only remaining nontrivial component of the field equations \eqref{eq:simplifiedFEQG} reads
\begin{align}
    R_{uu} - \Lambda g_{uu} - 4\kappa B_{uu} = -\frac{3}{2r^2}\Delta_{\mathbb H}^2 H = 0,
\end{align}
so that the metric \eqref{eq:4DMetricHyperbolicInteresting0} is an exact solution if and only if $H$ satisfies the biharmonic equation
\begin{align}\label{eq:PDEonH1}
    \boxed{\Delta_{\mathbb H}^2 H=0.}
\end{align}

\noindent Formally, this is exactly the same equation \eqref{eq:PDEonH0} as the one we obtained in the case of spherical wave surfaces. The solutions are again precisely the biharmonic functions, in this case on hyperbolic space. Again we note that any harmonic function, i.e. $\Delta_{\mathbb H} H=0$, is automatically biharmonic, which, in contrast to the spherical case, already provides a large range of solutions that are defined on the entire hyperbolic plane. The class of biharmonic functions on hyperbolic space is strictly larger, though, as we discuss below.\\

\subsubsection{The general solution to $\Delta_{\mathbb H}^2 H=0$}

\noindent  While above we have employed the Poincar\'e half-space model for hyperbolic space, harmonic solutions on $\mathbb H^2$ can be analyzed most easily in the Poincaré disk model of the hyperbolic plane. The relation between the half-space model and the disk model is provided by the Cayley transform. Denoting a point $(x,y)$ in the upper half plane by the corresponding complex number $z=x+iy$, the Cayley transform can be expressed as
 \begin{align}
    z \mapsto \frac{z - i}{z+i}.
 \end{align}
 This map is a global diffeomorphism between the upper half plane and the unit disk, and in the disk coordinates the hyperbolic metric is given by
\begin{align}\label{eq:hyperbolic_disk}
    \D s^2 = \frac{4r^2(\D x^2 + \D y^2)}{\left(1-(x^2 + y^2)\right)^2},\qquad \sqrt{x^2+y^2}<1
\end{align}
and the Laplace-Beltrami operator takes the form
\begin{align}
    \Delta_{\mathbb H} H= f(x,y)\Delta_{\mathbb E} H,
\end{align}
where $\Delta_{\mathbb E} = \partial_x^2  + \partial_y^2$ is the Euclidean Laplace-Beltrami operator, and 
\begin{align}
    f(x,y)=\tfrac{1}{4r^2}\left(1-(x^2 + y^2)\right)^2\neq 0.
\end{align}
Harmonic functions on $\mathbb H^2$ thus coincide, in these coordinates, precisely with the harmonic functions on the open unit disk in $\mathbb R^2$. Now consider the equation $\Delta_{\mathbb H}^2 H=0$. It is clearly equivalent to the pair of equations
\begin{align}
    \Delta_{\mathbb H}H = \phi, \qquad \Delta_{\mathbb H}\phi = 0,
\end{align}
which in turn is equivalent, using the form of $\Delta _{\mathbb H}$, to
\begin{align}
    \Delta_{\mathbb E}H &= \frac{\phi}{f}, \label{eq:effective_Poisson_eq}\\
    \Delta_{\mathbb E}\phi &= 0.\label{eq:effective_Laplace_eq}
\end{align}
This pair of equations can be solved explicitly, given suitable boundary conditions. The general solution to \eqref{eq:effective_Laplace_eq} (assuming it is bounded) is given by the Poisson integral formula for the unit disk (see e.g. \cite{griffiths1989geometric}), i.e.
\begin{align}\label{eq:Poisson_formula}
    \phi(\rho,\phi) = \frac{1}{2\pi}\int_0^{2\pi} \frac{1-\rho^2}{\rho^2 +1 - 2\rho\cos(\phi-\phi_0)}\Phi(\phi_0)\D\phi_0
\end{align}
in polar coordinates $(x,y) = (\rho \cos\phi, \rho\sin\phi)$, and with $\Phi\in L^\infty(S^1)$. In particular, $\phi$ is uniquely determined by its values on the unit circle $S^1$. 

Once $\phi$ is determined, the solution to the inhomogeneous equation \eqref{eq:effective_Poisson_eq} can be constructed using the Green's function for the (Euclidean) Poisson equation in the unit disk (see e.g \cite{courant2008methods}), which reads
\begin{align}
    G(\rho,\phi,\rho_0,&\phi_0) =  \nonumber\\
    &\frac{1}{4\pi}\log \left(\frac{\rho^2+\rho_0^2 - 2\rho\rho_0\cos(\phi-\phi_0)}{\rho_0^2\rho^2 +1 - 2\rho_0\rho\cos(\phi-\phi_0)}\right) .
\end{align}
The general solution to $\Delta_{\mathbb H}^2 H =0$ can then be expressed in (polar) Poincar\'e disk coordinates as
\begin{align}
    H(\rho ,&\phi) = \!\!\int_0^{2\pi}\!\!\!\int_0^1 \!G(\rho,\phi,\rho_0,\phi_0) \frac{\phi(\rho_0,\phi_0)}{f(\rho_0,\phi_0)} \rho_0\D\rho_0\D\phi_0 \label{eq:General_sol_hyp_part1}\\%
    &+ \frac{1}{2\pi}\int_0^{2\pi} \frac{1-\rho^2}{\rho^2 +1 - 2\rho\cos(\phi-\phi_0)}h(\phi_0)\D\phi_0, \label{eq:General_sol_hyp_part2}
\end{align}
where $\phi$ is given by \eqref{eq:Poisson_formula} and where $h\in L^\infty(S^1)$.

Note that the second integral \eqref{eq:General_sol_hyp_part2} always exists, since the factor multiplying $h(\phi_0)$ in the integrand is $L^1$, while $h$ is $L^\infty$, and the same is true for the integral \eqref{eq:Poisson_formula} that defines $\phi$. For the first integral \eqref{eq:General_sol_hyp_part1}, however, the $1/(1-\rho^2)^{2}$ behavior of $1/f$ complicates things. A sufficient condition for the integral to exist is that $\rho\phi(\rho,\phi)/f(\rho,\phi)$ is bounded on the unit disk, but we have not been able to find a simple condition on $\phi$ alone that enforces such behavior. In fact, we expect nonvanishing functions $\phi$ satisfying said sufficient condition to be rare, which would indicate that most (but certainly not all, as shown below) solutions to $\Delta_{\mathbb H}^2H=0$ are in fact harmonic, $\Delta_{\mathbb H}H=0$. Such solutions can be obtained from the general solution above simply by setting $\phi=0$ (equivalently, $\Phi=0$).

\subsubsection{\label{sec:solving2} A simple example}

\noindent In order to give a simple example we return to the formulation in terms of the Poincar\'e half plane as in section \ref{sec:hyp_field_eqs} and consider the function
\begin{align}
    H(x) = a + bx + c\ln x + d \,x\ln x, \quad a,b,c,d\in\R.
\end{align}
This function is biharmonic, i.e. it satisfies \eqref{eq:PDEonH1}, and hence it provides a solution to the field equations. Moreover, we have
\begin{align}
    \Delta_{\mathbb H}H = -c + d\, x,
\end{align}
so as long as $c,d$ do not both vanish, this function is even properly biharmonic (i.e. biharmonic but not harmonic). Other proper biharmonic functions on hyperbolic space have been constructed explicitly e.g. in \cite{GUDMUNDSSON2018244}.\\

\SH{We end this section with a final remark about the reduction of the field equations to 
\begin{align}
   \left(2 \kappa \Delta_{\mathbb E}^2 - \Delta_{\mathbb E}\right) H =0, \quad \Delta_{\mathbb S}^2 H = 0, \quad\Delta_{\mathbb H}^2 H=0,
\end{align}
in the case of flat, spherical, and hyperbolic wave surfaces, respectively. 
We have discussed this on a case-by-case basis. Alternatively, the result for all three cases can be proven simultaneously by choosing coordinates on the wave surfaces corresponding to the metric
\begin{align}
    \D s^2 =\frac{d\rho^2}{1- k \rho^2} + \rho^2 d\theta^2.
\end{align}
This covers the flat case ($k=0$), the spherical case $(k=1)$, and the hyperbolic case ($k=-1)$. 
}


\section{Discussion}

\noindent In this paper, we have introduced two new classes of exact vacuum solutions of pp-wave type to quadratic gravity. The solutions have a constant Ricci scalar but are not Einstein and therefore they are not vacuum solutions to Einstein's field equations with or without cosmological constant. While Einstein's equations dictate that vacuum pp-waves in general relativity have flat wave surfaces when expressed in standard form, and can only exist in the absence of a cosmological constant, our solutions are pp-waves with curved wave surfaces of any nonvanishing constant curvature, which is linearly related to the (nonvanishing) value of the cosmological constant. This leads to two distinct possibilities. 

First, (i) if the cosmological constant is positive then the wave surfaces must be spherical. In this case we have shown that the solutions should be understood only locally; indeed if the wave surfaces are assumed to be globally spherical then only a single rather trivial solution exists, by virtue of the fact that there are no nonconstant smooth biharmonic functions on the sphere. 

Second, (ii) in the presence of a negative cosmological constant the wave surfaces must have a hyperbolic geometry. In this case, nontrivial solutions do exist globally and we have obtained the expression for the general solution. 

Finally, for a vanishing cosmological constant, the wave surfaces are necessarily flat and we recover the QG version of the Peres waves by Madsen, which contains in particular all GR vacuum pp-waves. 

As mentioned before, our approach has in some sense been complementary to the approach in  \cite{gyratonic_pp_waves_quad_grav}, in which nonvanishing gyratonic terms in the pp-wave metric were considered, while leaving the wave surfaces flat---in contrast to the approach we have taken here, generalizing the wave surfaces by allowing them to be curved, while assuming instead that the gyratonic terms all vanish. In future work, it would be interesting to investigate the scenario in which both gyratonic terms and curved wave surfaces are present. Similarly, it would be very much of interest to see whether the results derived in the present work can be generalized to wave surfaces of arbitrary---not necessarily constant---curvature.


\begin{acknowledgments}
\noindent A. Fuster would like to thank V. Pravda, A. Pravdova, and J. Podolsky for correspondence on the topic of this work.

\end{acknowledgments}

\appendix
\section{Addendum on the spherical harmonic decomposition}\label{app:spherical_harmonics_expansion}

\noindent Let $f$ be a smooth (i.e. infinitely differentiable) real-valued function on the sphere, and write
\begin{align}
     f = \sum_{\ell=0}^\infty\sum_{m=-\ell}^\ell f_{\ell m} Y_{\ell m}.
\end{align}
We will sketch the proof that the Laplace-Beltrami operator $\Delta_{\mathbb S} $ acts `term by term', i.e. 
\begin{align}
    \Delta_{\mathbb S}f = \sum_{\ell=0}^\infty\sum_{m=-\ell}^\ell f_{\ell m} \Delta_{\mathbb S} Y_{\ell m},
\end{align}
justifying the term-by-term differentiation in Eq. \eqref{eq:term-by-term_differentiation}. Even though this is often taken for granted when constructing solutions to a Poisson equation, we have never encountered a concise proof in the literature, so we think it is worth providing it here.

To simplify notation, condense the subscripts $\ell,m$ on $Y_{\ell m}$ into a single subscript $i$, denote the corresponding eignenvalues by $\lambda_i$, and denote the $L^2$ inner product between two functions $f,g$ by $\langle f,g\rangle$.

\begin{proof}
Let $f\in C^\infty(\mathbb S^2)\subset L^2(\mathbb S^2)$ and write $f = \sum a_iY_i$ with $a_i\in\mathbb R$, so that $a_j$ is given by $a_j=\langle f, Y_j\rangle$. By smoothness of $f$ we also have $\Delta_{\mathbb S}f\in C^\infty(\mathbb S^2)\subset L^2(\mathbb S^2)$ and hence $\Delta_{\mathbb S}f$ can be written as $\Delta_{\mathbb S}f = \sum b_iY_i$ with $b_i\in\mathbb R$, where coefficient $b_j$ is given by $b_j=\langle\Delta_{\mathbb S}f, Y_j\rangle$. The crucial property of $\Delta_{\mathbb S}$ is that it is a symmetric operator on $C^\infty(\mathbb S^2)$ (this follows from Green's second identity and the fact that $\partial(\mathbb S^2)=\emptyset$), i.e. since both $f$ and each $Y_j$ are smooth, we have
\begin{align}
b_j&=\langle\Delta_{\mathbb S}f, Y_j\rangle = \langle f, \Delta_{\mathbb S} Y_j\rangle = \langle f, \lambda_j Y_j\rangle  \\
&= \lambda_j \langle f,  Y_j\rangle =  \lambda_j a_j.
\end{align}
This shows that $\Delta_{\mathbb S}f = \Delta_{\mathbb S}(\sum a_iY_i) = \sum \lambda_i a_i Y_i =  \sum a_i \Delta_{\mathbb S} Y_i$, as desired.
\end{proof}


\bibliography{references}

\end{document}